\documentclass[fleqn,usenatbib]{mnras}

\usepackage{newtxtext,newtxmath}

\usepackage[T1]{fontenc}
\usepackage{ae,aecompl}

\usepackage{graphicx}	
\usepackage{amsmath}	
\usepackage{amssymb}	
\usepackage{multirow}
\usepackage{lipsum}
\usepackage{wasysym}

\newcommand{\ms}{m\,s$^{-1}$}
\newcommand{\kms}{km\,s$^{-1}$}
\newcommand{\mpl}{$M_{\rm{p}}$}

\newcommand{\mn}{$M_{\rm{\neptune}}$}
\newcommand{\vr}{V_{r}}

\newcommand{\prot}{$P_{\rm{rot}}$}

\newcommand{\Yk}{$\boldsymbol{Y_{\rm{k}}}$}
\newcommand{\Ys}{$\boldsymbol{Y_{\rm{s}}}$}
\newcommand{\Vs}{$\boldsymbol{V_{\rm{s}}}$}
\newcommand{\Km}{$K_{\rm{m}}$}
\newcommand{\Kp}{$K_{\rm{p}}$}
\newcommand{\Pp}{$P_{\rm{orb}}$}
\newcommand{\Php}{$\phi_{\rm{m}}$}
\newcommand{\Phii}{$\phi_{\rm{p}}$}
\newcommand{\sn}{$\sigma_{\rm{n}}$}

\newcommand{\Tc}{$\tau_{\rm{c}}$}
\newcommand{\Tb}{$\tau_{\rm{b}}$}
\newcommand{\hyp}{$\boldsymbol{\theta}$}
\newcommand{\hypb}{$\boldsymbol{\theta_{\rm{b}}}$}
\newcommand{\BF}{$B_{\rm{F}}$}

\defcitealias{david2016}{D16}
\defcitealias{mann2016}{M16}
\defcitealias{klein2019}{K19}

\title[Simulated mass measurement of K2-33b]{Simulated mass measurements of the young planet K2-33b}

\author[B. Klein et al.]{
Baptiste Klein,$^{1,2}$\thanks{E-mail: baptiste.klein@irap.omp.eu}
J.-F. Donati,$^{1,2}$
\\
$^{1}$Universite de Toulouse, UPS-OMP, IRAP, 14 Avenue E. Belin, Toulouse F-31400, France\\
$^{2}$CNRS, IRAP/UMR 5277, Toulouse, 14 Avenue E. Belin, Toulouse F-31400, France\\
}

\date{Accepted XXX. Received YYY; in original form ZZZ}

\pubyear{2019}

\begin{document}
\label{firstpage}
\pagerange{\pageref{firstpage}--\pageref{lastpage}}
\maketitle

\begin{abstract}
In this paper, we carry out simulations of radial velocity (RV) measurements of the mass of the 8-11~Myr Neptune-sized planet K2-33b using high-precision near infrared velocimeters like SPIRou at the Canada-France-Hawaii Telescope. We generate a RV curve containing a planet signature and a realistic stellar activity signal, computed for a central wavelength of 1.8~$\mu$m and statistically compatible with the light-curve obtained with K2. The modelled activity signal includes the effect of time-evolving dark and bright surface features hosting a 2~kG radial magnetic field, resulting in a RV signal of semi-amplitude $\sim$30~\ms. Assuming a 3-month visibility window, we build RV time-series including Gaussian white noise from which we retrieve the planet mass while filtering the stellar activity signal using Gaussian Process Regression. We find that 35/50 visits spread over 3 consecutive bright-time runs on K2-33 allow one to reliably detect planet RV signatures of respectively 10 and 5~\ms\ at precisions~$>3\sigma$. We also show that 30 visits may end up being insufficient in some cases to provide a good coverage of the stellar rotation cycle, with the result that the planet signature can go undetected or the mass estimation be plagued by large errors.


\end{abstract}




\section{Introduction}

Planet formation and evolution models critically need observational constraints on how planet bulk densities vary with time in the early stages of their lives \citep[e.g.][]{mordasini2012,alibert2013}. This requires to measure radii of transiting planets through the relative depths of their photometric transits on the one hand, and masses through the semi-amplitudes of their radial velocity (RV) curves, on the other hand. Both measurements are challenging for pre-main-sequence (PMS) stars known to exhibit intense magnetic activity \citep[e.g.][]{bouvier1989} inducing photometric and RV fluctuations that largely overshadow the planet signatures \citep[e.g.][]{crockett2012}. As a result, only a handful of candidate close-in giant planets younger than 20~Myr have been unveiled so far, either using RV observations
\citep[][]{donati2016,johns-krull2016J,yu2017} or transit photometry \citep{david2016,mann2016,david2019,david2019b}. None of them have a well-measured bulk density.

Observing PMS stars in the near infrared (nIR) rather than in the V band should make it easier to separate the planet signature from the stellar activity signal as the latter is expected to be weaker in this domain \citep{mahmud2011} and the stars are significantly reddened. High-precision nIR velocimeters like SPIRou \citep{donati2018}, CARMENES \citep{Quirrenbach2014}, GIARPS \citep{claudi2017} or NIRPS \citep{wildi2017} are thus the most promising instruments worldwide to carry out mass measurements of close-in transiting planets orbiting PMS stars. Magnetic fields are however expected to affect stellar RV activity signals \citep[][]{reiners2013,hebrard2014}, making the problem non trivial and worth a detailed simulation study.  This is especially relevant given that 300 nights of CFHT time are already allocated to the SPIRou Legacy Survey (SLS), some of them being dedicated to the RV follow-up of stars hosting transiting planets, with the goal of measuring the planet masses. 


K2-33 is a 8-11~Myr M3 PMS star located in Upper Scorpius that was shown to host a 5-$R_{\rm{\oplus}}$ close-in transiting planet \citep[][hereafter \citetalias{david2016} and \citetalias{mann2016} respectively]{david2016,mann2016} from the 80-d continuous light-curve obtained during campaign 2 of the K2 mission \citep{howell2014}. K2-33 will be observed with SPIRou as part of the SLS in an attempt to measure the mass of its close-in planet through RV observations. In this study, we propose to use K2-33 as a representative of the PMS stars to be observed within the SLS. We simulate SPIRou RV observations of this star and attempt retrieving the RV signature of the Neptune-sized planet assuming various planet masses, sampling schemes and levels of white noise. In Sec.~\ref{sec:Section2}, we outline how we generate the realistic synthetic time-series for K2-33 and, in Sec.~\ref{sec:section3}, their modelling in order to filter the stellar activity signal while estimating the planet parameters. We finally summarise our results and discuss their implications in Sec.~\ref{sec:section5}.

\section{Synthetic RV data sets}\label{sec:Section2}

\begin{table*}
    \centering
    \caption{Prior densities ($\uppi$; columns 2 and 3), and best estimates of \hyp\ (columns 4 to 6) obtained when modelling the stellar activity photometry (\Yk\ and \Ys\ for K2 and synthetic curves respectively) and RV (\Vs) curves. $\mathcal{U}$ stands for the uniform distribution. Note that we use the Gaussian distribution $\mathcal{N}(6.35 d,0.04 d)$ as a prior density law for $\theta_{3}$ when modelling the RV time-series described in Sec~\ref{sec:section3}.}
    \label{tab:GP_LC}
    \begin{tabular}{cccccc}
        \textbf{Param.} & $\uppi \left( \boldsymbol{\theta} | \boldsymbol{Y_{\rm{k}}} \right)$, $\uppi \left( \boldsymbol{\theta} | \boldsymbol{Y_{\rm{s}}} \right)$ & $\uppi \left( \boldsymbol{\theta} | \boldsymbol{V_{\rm{s}}} \right)$ & \Yk & \Ys  & \Vs \\
        \hline
        $\ln \theta_{1} $ & $\mathcal{U}(-10,1)$ & $\mathcal{U}(-2,7)$ &  $-4.6 \pm 0.2 $  & $-4.9 \pm 0.2$  & 2.9 $\pm$ 0.2 [$\ln$ \ms] \\
        $ \ln \left( \theta_{2} \text{ [d]} \right) $  & $\mathcal{U}(1.0,7.0)$ & $\mathcal{U}(1.0,7.0)$ & 2.94 $\pm$ 0.04  & 3.27 $\pm$ 0.08 & 3.58 $\pm$ 0.10 \\
        $\theta_{3}$ [d]  & $\mathcal{U}(5.5,7.5)$ & $\mathcal{U}(6.25,6.65)$ & 6.35 $\pm$ 0.04 & 6.36 $\pm$ 0.02 & 6.35 $\pm$ 0.01  \\
        $\theta_{4}$ & $\mathcal{U}(0.1,5.0)$ & $\mathcal{U}(0.1,4.0)$ &  1.01 $\pm$ 0.1  & 0.71 $\pm$ 0.06 & 0.33 $\pm$ 0.02 \\
        \hline
    \end{tabular}
\end{table*}



Our method to simulate RV observations of K2-33 with SPIRou is similar to that described in  \citet[][hereafter \citetalias{klein2019}]{klein2019}. We first generate a densely-sampled RV curve, containing a planet signature and a stellar activity signal whose statistical properties are consistent with that of the K2 light curve which encloses information about the evolution properties of the features at the surface of the star. We then create RV time-series by selecting observation dates using different schemes and adding noise to account for the various sources that may affect the data.

\subsection{Stellar activity RV curve}

Using the method described in \citetalias{klein2019}, we model the stellar surface into a dense grid of 100 000 cells and generate 80~d densely-sampled photometric and RV curves. This model includes time-evolving bright/dark features whose effective temperatures are scaled from \citet{berdyugina2005} and whose brightnesses are inferred using Planck's law at K2 and SPIRou central wavelengths respectively for the photometric and RV curves. Each feature is also assumed to host a radial magnetic field of 2~kG to account for the significant Zeeman splitting of the line profiles at SPIRou's central wavelength \citep[e.g.][]{reiners2013}.

\begin{figure}
    \centering
    \includegraphics[width=\linewidth]{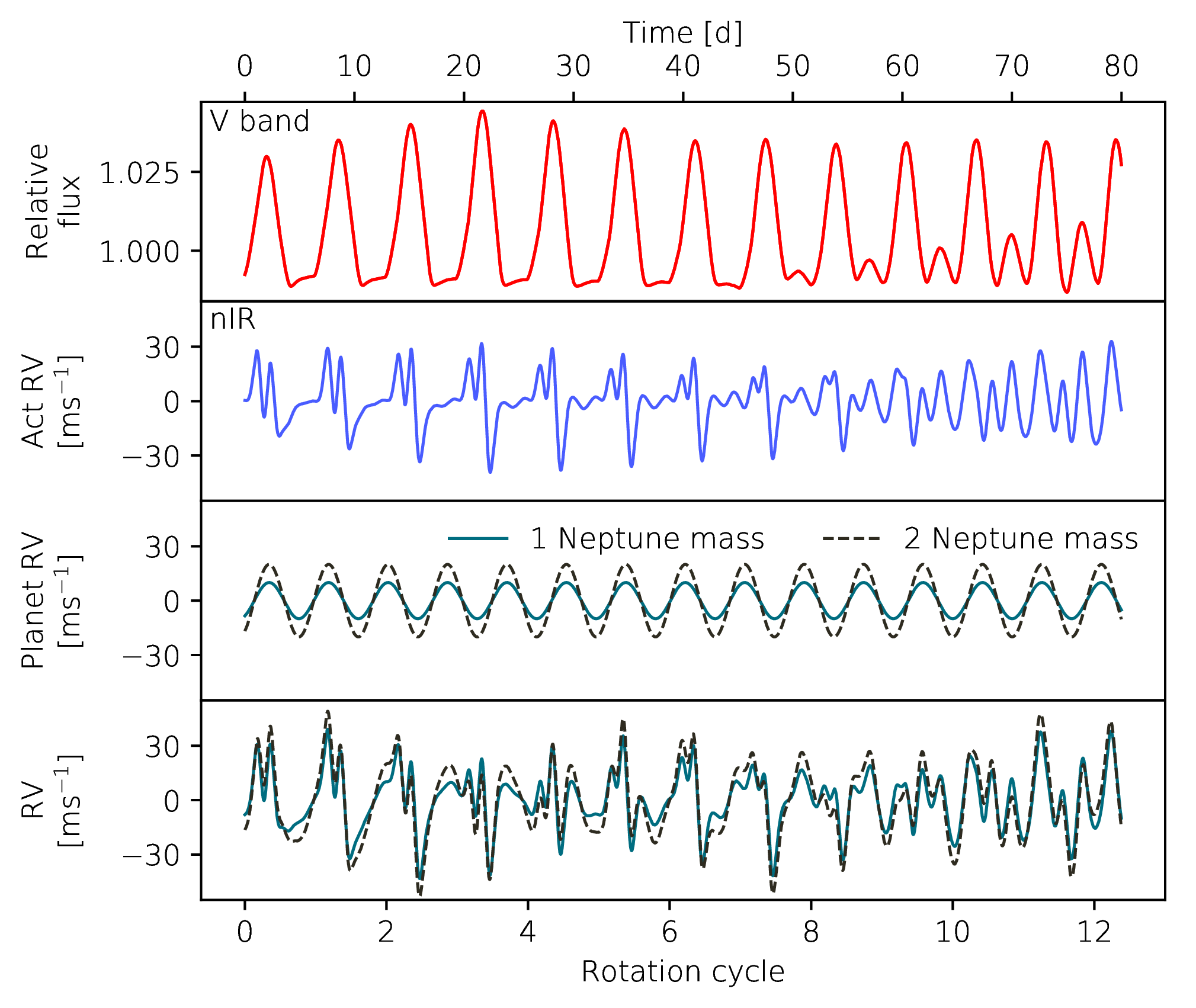}
    \caption{Synthetic photometry and RV curves of K2-33. From top to bottom: synthetic relative brightness curve in K2 spectral band; corresponding stellar activity RV curve at SPIRou's central wavelength; expected planet RV signatures for semi-amplitudes of 10 and 20~\ms\ (blue solid line and black dashed line respectively); resulting RV curves.}
    \label{fig:sig_tot}
\end{figure}

\begin{figure}
    \centering
    \includegraphics[width=\linewidth,height=4.5cm]{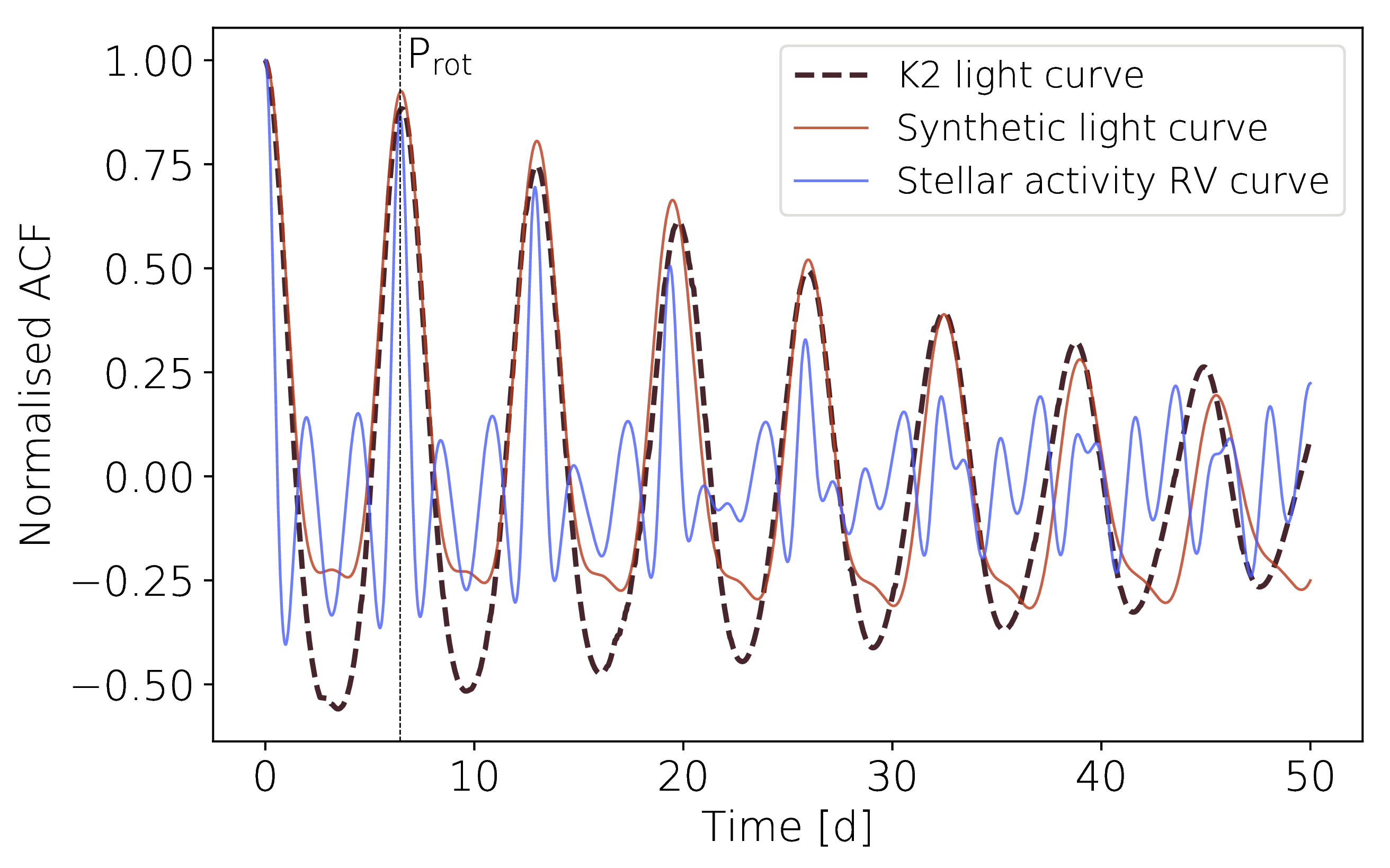}
    \caption{ACFs of the \textsc{everest}-detrended K2 light-curve (black dashed line), synthetic photometric curve (red solid line) and synthetic stellar activity RV curve (blue solid line). The stellar rotation period is indicated by the vertical dashed line.}
    \label{fig:acf}
\end{figure}

We tune the appearance probability $p_{\rm{r}}$ of activity features at the surface of the star at each time step as well as their lifetimes $t_{\rm{r}}$ and maximum relative areas $s_{\rm{r}}$ so as the autocorrelation function (ACF) of the synthetic photometric curve is similar to that of the K2 light-curve detrended with the \textsc{everest} software \citep{luger2016,luger2018}. The stellar parameters adopted to generate the stellar activity curves are shown in Table~\ref{tab:stellar_prop}. Note that we rescaled the total lifetime of each injected feature by a factor $\propto s_{\rm{r}}$ so that the larger the feature, the longer its lifetime, in agreement with what is observed on the K2 light-curve. The photometry and RV curves resulting from our activity model are shown in Fig.~\ref{fig:sig_tot}. Their ACFs are superimposed to that of the detrended K2 light-curve in Fig.~\ref{fig:acf}, showing a good overall agreement.


We then compare the statistical properties of the newly-synthesized photometric and RV curves to that of the K2 light-curve by independently modelling the rotationally-modulated component within each curve using Gaussian Process Regression \citep[GPR;][]{rasmussen2006}. The K2 light-curve being affected by short-lived phenomena like flares and planetary transits, we use a specific process to extract the signal produced by inhomogeneities at the surface of the star. We first build a reduced photometric data set by dividing the K2 light-curve into 200 consecutive time intervals of equal duration and then by computing the median of all the points within each interval. We then model the resulting 200-point light-curve using GPR assuming a quasi-periodic covariance kernel \citep[e.g.][]{haywood2014}, $k$, which relies on a vector of 4 so-called hyperparameters, \hyp, such that

\begin{eqnarray}
k(t_{i},t_{j}) = \theta_{1}^{2} \exp \left[ - \frac{(t_{i}-t_{j})^{2}}{\theta_{2}^{2}} - \frac{\sin^{2} \frac{\uppi (t_{i}-t_{j})}{\theta_{3}}}{\theta_{4}^{2}} \right],
\label{eq:cov_fct}
\end{eqnarray}

\noindent
where $t_{i}$ and $t_{j}$ are the times associated with observations $i$ and $j$, and $\theta_{1}$ to $\theta_{4}$ respectively stand for the amplitude, evolution timescale, recurrence period and smoothing factor of the Gaussian Process (GP). \hyp\ is estimated by maximising its posterior posterior density sampled using a Bayesian Markov Chain Monte Carlo (MCMC) process. We then use the trained GP to predict the values of the rotationally-modulated photometric signal at the times of K2 observations and reject the points deviating from the GPR prediction by more than $3 \sigma$, then repeat the process until no point is rejected. The prior densities used for the hyperparameters as well as their best estimates are shown in Table~\ref{tab:GP_LC}.

We select 250~evenly sampled data points from the densely-sampled synthetic photometric and RV curves, add a Gaussian white noise of respectively $\sim$300~ppm and 2~\ms\ and use GPR to independently model the two data sets (called \Ys\ and \Vs\ respectively) in order to estimate their statistical properties. The outcome of the fit as well as the adopted prior densities are given in Table~\ref{tab:GP_LC} (rms of the residuals of 300~ppm and 1.8~\ms\ for \Ys\ and \Vs\ respectively). The significantly larger $\theta_{2}$ in the RV time-series comes from the fact that the RV activity signal is dominated by the larger, slowly-evolving features, whereas the signatures of the smaller, rapidly-evolving features are partly drowned in the Gaussian white noise. Conversely, we note that $\theta_{4}$ is roughly twice as large in photometry than in RV, which is expected given that a feature at the stellar surface produces a RV signature that evolves roughly twice as fast as its photometric counterpart \citep[see][]{aigrain2012}. We also observe notable variations of $\theta_{2}$ and $\theta_{4}$ between the 2 photometric time-series (\Yk\ and \Ys), indicating (i)~slightly longer evolution timescales for the modelled features and (ii)~a lower smoothing factor for the synthetic light-curve.

\begin{table}
    \centering
    \caption{Main stellar properties used to generate stellar activity photometric and RV curves for K2-33.}
    \label{tab:stellar_prop}
    \begin{tabular}{c|c|c}
        \textbf{Parameter} & \textbf{Value} & \textbf{Notes} \\
        \hline
        $P_{\rm{rot}}$ & 6.35~d & Main peak in ACF (see Fig.~\ref{fig:acf}) \\
        $v \sin{i}$ & 8.2~\kms & From \citetalias{mann2016} \\
        $T_{\rm{eff}}$ & 3475~K & Average between \citetalias{mann2016} and \citetalias{david2016} \\
        $i$ & 88$\degr$ & Average between \citetalias{mann2016} and \citetalias{david2016} \\
        $p_{\rm{r}}$ & 0.3~\% & -- \\
        $\log_{10} s_{\rm{r}}$ & $\mathcal{N}\left(-2.6,0.1\right)$ & -- \\
        $t_{\rm{r}}$ & 18.0~\prot & Rescaled by a factor  $\propto s_{\rm{r}}$ \\
        \hline
    \end{tabular}
\end{table}

\subsection{Planet RV curve}


Consistently with the modelled planet transit curves of \citetalias{mann2016}, we assume that K2-33b's orbit is circular, and thus that the expected RV signal that the planet induces in the spectrum of its host star is given by


\begin{eqnarray}
\vr(t) = K_{\rm{p}} \cos{ \left( \frac{2 \uppi}{P_{\rm{orb}}} t+\phi_{\rm{p}} \right) },
\label{eq:RV_circ}
\end{eqnarray}

\noindent
where \Kp, \Pp\ and \Phii\ are respectively the semi-amplitude, orbital period and phase of the signal. Since the planet mass is the main parameter we aim at characterizing, and as such is not well known, we consider different values for \Kp\ listed along with the corresponding planet mass (computed using the parameters reported in \citetalias{david2016} or \citetalias{mann2016}) in Table~\ref{tab:pl_param}. In what follows, we assume that the planet orbital period and phase are given by the average of the values measured by \citetalias{david2016} and \citetalias{mann2016}.

\begin{table}
    \centering
     \caption{Summary of K2-33b parameters reported in \citetalias{david2016} and \citetalias{mann2016} (columns 3 and 4 respectively). The planet masses, \mpl, are computed by means of Kepler's third law using the stellar mass and planet orbital period reported in each article (\mn\ standing for Neptune mass). }
    \label{tab:pl_param}
    \begin{tabular}{c|c|c|c}
    \hline
    Source &  &  \citet{david2016} & \citet{mann2016} \\
    \hline
     \multirow{3}{*}{ \Kp }  & 5~\ms &  \mpl\ = 0.37~\mn & \mpl\ = 0.54~\mn \\
     & 10~\ms & \mpl\ = 0.73~\mn & \mpl\ = 1.09~\mn \\
     & 20~\ms   &  \mpl\ = 1.47~\mn  & \mpl\ = 2.18~\mn \\
     \hline
     \Pp &  & 5.42513~d & 5.424865~d \\
     $\phi_{\rm{p}}$ &  & 0.374~rad & 0.375~rad \\
     \hline
    \end{tabular}
\end{table}

\subsection{Creating the time-series}\label{sec:section2.3}

From the densely-sampled stellar activity and planet curves described above, we generate RV time-series assuming that observations are carried out from the CFHT during 3 consecutive 15d-bright time periods centered on full moons. Among the potentially observable nights included in our visibility window, we build data sets with $N$~=~30, 35, 40 randomly-selected data points at airmass \la1.8. We also consider a more optimistic case with data sets of $N$~=~50 data points randomly-selected on 3 consecutive CFHT bright time periods of 20~d each. For each value of $N$, we build 8 sets of 10 RV time-series with different realizations of white noise, each set being drawn using a different distribution of observing epochs. We deal with the various sources of noise expected to pollute the RV time-series (e.g. photon and instrument noises) by adding a centered Gaussian white noise with standard deviations, \sn, of 2~\ms\, for an optimistic case, or 5~\ms\ for a more conservative case achievable for velocimeters like SPIRou on as faint a target as K2-33.

\section{Modelling the time-series}\label{sec:section3}

\subsection{Method}\label{sec:3.1}

We model the synthetic RV time-series using the method detailed in \citetalias{klein2019}. Assuming that \Pp\ is known from photometry, the planet RV signature described by Eq.~\ref{eq:RV_circ} can be expressed as a linear function of parameters depending on \Km\ and \Php, i.e. respectively the semi-amplitude and orbital phase of the RV signal induced by the planet on the stellar spectrum to recover. The stellar activity RV signal is modelled using GPR assuming the 4-parameter quasi-periodic covariance kernel defined in Eq.~\ref{eq:cov_fct}. We use the \textsc{emcee} affine invariant sampler \citep[][5000 iterations of 100 walkers]{Foreman-Mackey2013} to sample the posterior density of the model marginalized over the planet parameters. The prior densities adopted for \hyp\ are given in Table~\ref{tab:GP_LC}. The median and 1$\sigma$ error bars on \hyp\ are computed from the posterior densities after removing a burn-in period of $\sim$100\,000 steps (i.e. 1000 iterations of 100 walkers), while the planet parameters are estimated using a least-square estimator computed at the value of \hyp\ that maximises the likelihood of the model (hereafter \hypb). The resulting Gaussian posterior density for \Km\ is convolved with the distribution of the semi-amplitudes of the planetary signal obtained by applying the aforementioned least-square estimator to $\sim$50\,000 samples \hyp\ (i.e. 500 iterations of 100 walkers), in order to account for correlations between planet and stellar activity parameters when computing the error bars. This process results in increasing the error bars on \Km\ by typically a few \%.

The quality of the planet detection within the RV time-series is assessed using the so-called Bayes factor \citep[\BF, see Eq.~2 from][for a proper definition]{diaz2014} to compare the marginal likelihoods of models assuming 0 (i.e. stellar activity and white noise only) and 1 planet signature in the data sets. The resulting \BF, computed using the method introduced in \citet{Chib2001}, allows to derive the posterior odds ratio which gives an estimation of the significance of the retrieved planet signal in each data set. Following \citet{Jeffreys1961}, we take \BF\ $>150$ (5 in log) to be the criterion to diagnose a fair detection of K2-33b.

\subsection{Results}

We independently model the mock RV time-series described in Sec.~\ref{sec:section2.3} using the process described in Sec.~\ref{sec:3.1} and reject those for which (i)~the MCMC process did not converge (e.g. posterior density presenting multiple local minima or stuck on one of the prior boundaries) or (ii)~\BF\ lies below the planet detection threshold.

The outcomes for each value of $N$ and \Kp, averaged over all the 8 sets of 10 RV time-series, are given in Tables~\ref{tab:results_npts_2ms} and~\ref{tab:results_npts_5ms} for \sn\ of 2 and 5~\ms\ respectively. We note that respectively 18-24~\% and 25-33~\% of data sets with $N = 30$ are rejected for the 2 cases of considered white noises, mainly due to the fact the MCMC process tends to minimize $\theta_{4}$, resulting in strongly over-fitting the data. Moreover, \Km\ is strongly over-estimated in the remaining RV time-series, consistently with \citet{damasso2019} for similar cases with too few observational constraints. \BF\ is also affected by this over-estimation, as evidenced by its surprisingly high values, especially for the lowest \Kp\ considered in this study. This shows that most data sets with $N = 30$ are too sparse to provide a dense-enough coverage of the rotation cycle, leading to excess flexibility for the GP and resulting in erroneous estimates for \Km. 

\begin{figure}
\centering
\includegraphics[width=\linewidth]{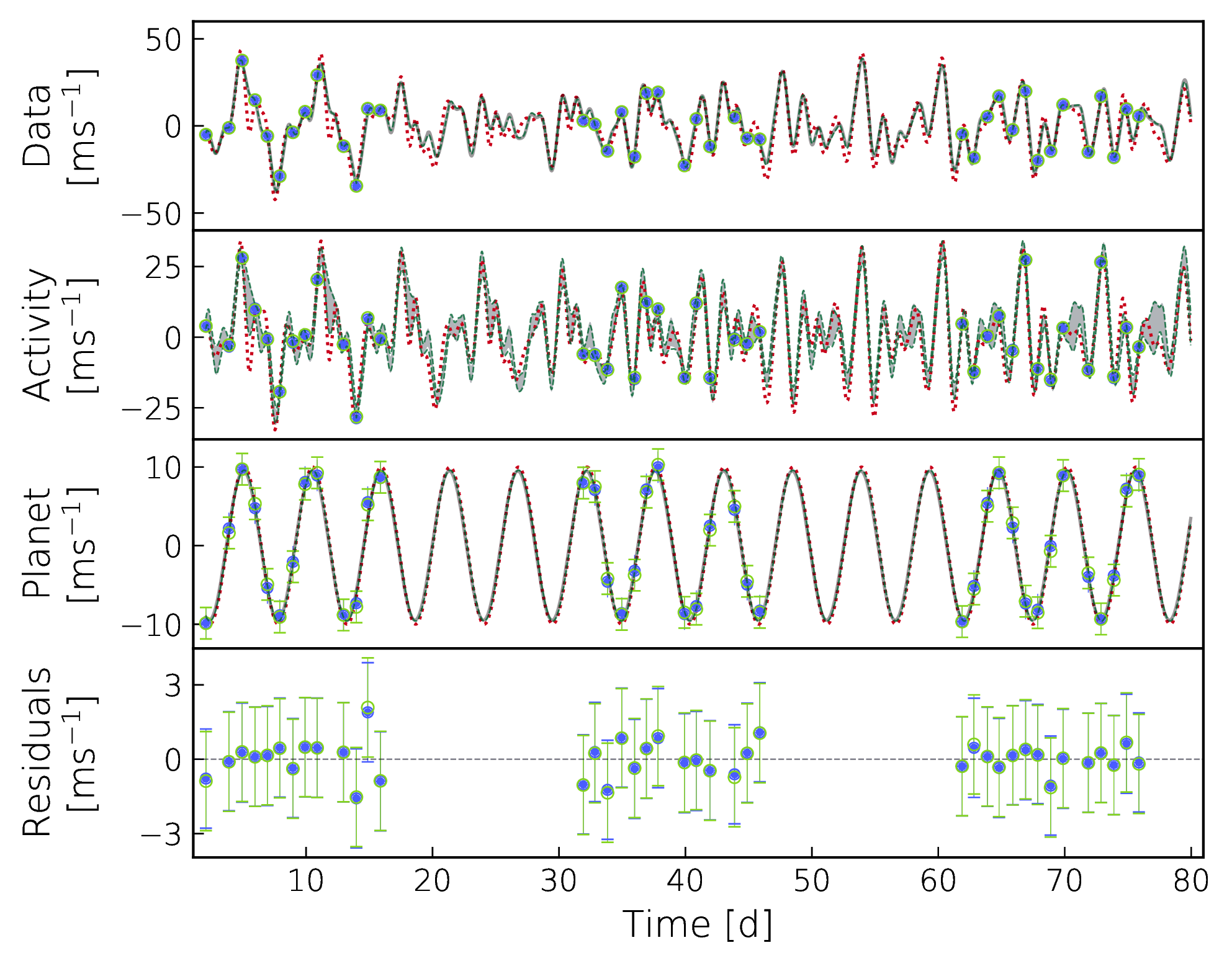}
\caption{Best fit to one of our synthetic RV data sets containing $N$~=~40 data points and assuming \Kp\ = 10~\ms\ and \sn\ = 2~\ms\ rms. From top to bottom: RV data set, stellar activity RV signal, planet signature and residuals. In each panel, the true RV signal is shown in red dotted lines, while the best predictions from the model are shown in green dashed lines and grey solid lines (resp. 1$\sigma$ error bands for panel 2), respectively when imposing the orbital phase of the planet RV signal to be that derived from the photometry and when regarding it as a free parameter. Curves and data points in the 2 middle panels are obtained by subtracting from the synthetic data all the modelled components except that displayed in the panel.}
\label{fig:pred_M12_free}
\end{figure}

The reconstruction of the data sets is considerably improved for $N\geq35$, as evidenced by (i)~the good agreement between the statistical properties of the RV curve (see Table~\ref{tab:GP_LC}) and the output GP parameters and (ii)~the more accurate retrieval of \Km\ (see also the illustration given in Fig.~\ref{fig:pred_M12_free}). The over-estimation noted for \Km\ at $N = 30$ strongly decreases at $N = 35$ and is no longer significant at $N$\ga40. We find that planet signatures of \Kp\ \ga~10~\ms\ are reliably recovered at precisions of 5 and 4$\sigma$ for \sn\ of 2 and 5~\ms\ respectively. Finally, \Kp\ =~5~\ms\ is fairly detected at 3$\sigma$ for $N = 50$ data points while it remains marginally recovered for lower number of visits.

Imposing the planet orbital phase to be that derived from photometry barely improves the precision of \Km\ and does not significantly impact the detectability of the planet. 
Moreover, we find that low-to-moderate elliptical planet signatures with eccentricities $\la$0.2 marginally impacts \Km\ as well as its error bars. We thus expect our algorithm to yield accurate estimates of the mass of K2-33b whose orbit is unlikely to be strongly elliptical \citep[see][]{mann2016}. Constraining the eccentricity of moderately elliptical planet orbits will require significantly more measurements that what we propose in this study, as assessed by preliminary simulations showing that uncertainties of the order of 0.15 on the eccentricity of the planet orbit require typically 100 visits to be achieved.

Finally, as already discussed in \citetalias{klein2019}, in the specific case where the amplitude of the stellar activity signal is significantly larger than \sn\ and the RV time-series contain a low number of data points, the GP tends to adapt by slightly adjusting \hypb\ to partly reconstruct the noise. Note that this trend decreases when \sn\ and/or $N$ increase in our simulations (see Tables~\ref{tab:results_npts_2ms} and \ref{tab:results_npts_5ms}).

\begin{table*}
\centering
\caption{Results of the fit to RV time-series for various $N$ and \Kp\ at \sn~=~2~\ms. \Tc\ and \Tb\ stand for the fractions of data sets rejected due to an unrealistic \hypb\ and a too low value for \BF\ respectively. The values shown in each line are averaged over all the remaining RV data sets of given $N$ and \Kp.}
\label{tab:results_npts_2ms}
\begin{tabular}{cccccccccccc}
\hline
\Kp & N$_{\rm{pts}}$ & \Tc & \Tb & $\ln \theta_{1}$ & $\ln \theta_{2}$ & $\theta_{3}$ & $\theta_{4}$ & \Km & rms & ln $\mathcal{L}_{\rm{max}}$ & $\ln$ BF\\

[\ms] & & [\%] & [\%] & [$\ln$ \ms] & [$\ln$ d] & [d] & & [\ms] & [\ms] & \\

\hline
5 & 30 & 23 & 1 & $2.52 \pm 0.18$ & $3.79 \pm 0.70$ & $6.37 \pm 0.03$ & $0.24 \pm 0.07$ & $8.9 \pm 2.7$ & 0.46 & $-106.9 \pm 0.9$ & $10.2 \pm 0.9$ \\ 
5 & 35 & 3 & 3 & $2.52 \pm 0.17$ & $3.71 \pm 0.55$ & $6.35 \pm 0.03$ & $0.23 \pm 0.06$ & $7.4 \pm 2.1$ & 0.53 & $-124.5 \pm 1.4$ & $9.1 \pm 1.0$ \\ 
5 & 40 & 0 & 3 & $2.56 \pm 0.17$ & $3.68 \pm 0.32$ & $6.36 \pm 0.02$ & $0.27 \pm 0.05$ & $5.9 \pm 1.8$ & 0.55 & $-140.8 \pm 1.6$ & $8.0 \pm 1.2$ \\ 
5 & 50 & 0 & 0 & $2.67 \pm 0.18$ & $3.59 \pm 0.17$ & $6.35 \pm 0.02$ & $0.27 \pm 0.04$ & $6.0 \pm 1.1$ & 0.64 & $-175.4 \pm 2.0$ & $13.9 \pm 1.9$ \\ 
\hline
10 & 30 & 21 & 1 & $2.51 \pm 0.18$ & $3.96 \pm 0.76$ & $6.38 \pm 0.03$ & $0.25 \pm 0.07$ & $13.5 \pm 2.6$ & 0.45 & $-107.2 \pm 1.2$ & $14.2 \pm 0.8$ \\ 
10 & 35 & 4 & 0 & $2.52 \pm 0.17$ & $3.70 \pm 0.53$ & $6.35 \pm 0.03$ & $0.23 \pm 0.06$ & $11.4 \pm 1.9$ & 0.53 & $-123.2 \pm 1.1$ & $14.9 \pm 1.1$ \\ 
10 & 40 & 0 & 0 & $2.55 \pm 0.17$ & $3.68 \pm 0.33$ & $6.36 \pm 0.02$ & $0.26 \pm 0.05$ & $10.6 \pm 1.7$ & 0.54 & $-141.3 \pm 1.5$ & $15.6 \pm 1.4$ \\ 
10 & 50 & 0 & 0 & $2.67 \pm 0.17$ & $3.59 \pm 0.17$ & $6.35 \pm 0.02$ & $0.28 \pm 0.04$ & $10.8 \pm 1.1$ & 0.66 & $-175.8 \pm 2.1$ & $23.0 \pm 1.7$ \\ 
\hline
20 & 30 & 18 & 0 & $2.52 \pm 0.18$ & $3.90 \pm 0.77$ & $6.38 \pm 0.03$ & $0.23 \pm 0.07$ & $23.2 \pm 2.6$ & 0.47 & $-106.7 \pm 1.1$ & $22.1 \pm 0.8$ \\ 
20 & 35 & 1 & 0 & $2.52 \pm 0.17$ & $3.66 \pm 0.54$ & $6.35 \pm 0.03$ & $0.23 \pm 0.06$ & $20.6 \pm 1.9$ & 0.53 & $-124.7 \pm 1.4$ & $25.3 \pm 1.4$ \\ 
20 & 40 & 0 & 0 & $2.56 \pm 0.17$ & $3.63 \pm 0.32$ & $6.36 \pm 0.03$ & $0.27 \pm 0.05$ & $20.4 \pm 1.8$ & 0.55 & $-141.3 \pm 1.7$ & $26.9 \pm 1.6$ \\ 
20 & 50 & 0 & 0 & $2.67 \pm 0.17$ & $3.59 \pm 0.17$ & $6.35 \pm 0.02$ & $0.27 \pm 0.04$ & $20.8 \pm 1.0$ & 0.65 & $-175.0 \pm 1.7$ & $37.0 \pm 1.4$ \\ 
\hline

\end{tabular} 
\end{table*}

\begin{table*}
\centering
\caption{Same as Table~\ref{tab:results_npts_2ms} for \sn\ =~5~\ms.}
\label{tab:results_npts_5ms}
\begin{tabular}{cccccccccccc}
\hline

\Kp & $N$ & \Tc & \Tb & $\ln \theta_{1}$ & $\ln \theta_{2}$ & $\theta_{3}$ & $\theta_{4}$ & \Km & rms & ln $\mathcal{L}_{\rm{max}}$ & $\ln$ BF \\

[\ms] & & [\%] & [\%] & [$\ln$ \ms] & [$\ln$ d] & [d] & & [\ms] & [\ms] & & \\

\hline
5 & 30 & 33 & 1 & $2.52 \pm 0.21$ & $3.76 \pm 0.95$ & $6.37 \pm 0.03$ & $0.25 \pm 0.09$ & $10.0 \pm 3.0$ & 2.27 & $-109.7 \pm 2.3$ & $8.5 \pm 1.4$ \\ 
5 & 35 & 9 & 1 & $2.50 \pm 0.20$ & $3.85 \pm 0.81$ & $6.36 \pm 0.03$ & $0.23 \pm 0.08$ & $8.5 \pm 2.5$ & 2.49 & $-128.2 \pm 2.1$ & $8.4 \pm 1.5$ \\ 
5 & 40 & 1 & 8 & $2.56 \pm 0.19$ & $3.75 \pm 0.50$ & $6.36 \pm 0.03$ & $0.25 \pm 0.06$ & $7.1 \pm 2.6$ & 2.53 & $-147.5 \pm 2.6$ & $7.1 \pm 1.4$ \\ 
5 & 50 & 1 & 0 & $2.60 \pm 0.18$ & $3.68 \pm 0.29$ & $6.36 \pm 0.02$ & $0.25 \pm 0.06$ & $6.3 \pm 1.6$ & 2.73 & $-184.7 \pm 2.7$ & $10.8 \pm 2.2$ \\ 
\hline
10 & 30 & 28 & 0 & $2.55 \pm 0.21$ & $3.88 \pm 0.90$ & $6.37 \pm 0.03$ & $0.25 \pm 0.08$ & $14.0 \pm 3.0$ & 2.32 & $-110.1 \pm 2.0$ & $12.3 \pm 1.4$ \\ 
10 & 35 & 10 & 0 & $2.52 \pm 0.20$ & $3.90 \pm 0.88$ & $6.36 \pm 0.03$ & $0.23 \pm 0.08$ & $12.1 \pm 2.6$ & 2.50 & $-128.6 \pm 2.5$ & $13.1 \pm 1.9$ \\ 
10 & 40 & 3 & 0 & $2.58 \pm 0.19$ & $3.83 \pm 0.54$ & $6.36 \pm 0.03$ & $0.25 \pm 0.06$ & $11.3 \pm 2.5$ & 2.55 & $-147.9 \pm 1.9$ & $12.6 \pm 1.6$ \\ 
10 & 50 & 0 & 0 & $2.62 \pm 0.18$ & $3.66 \pm 0.30$ & $6.36 \pm 0.02$ & $0.24 \pm 0.05$ & $11.6 \pm 1.6$ & 2.69 & $-185.5 \pm 2.4$ & $18.7 \pm 2.0$ \\ 
\hline
20 & 30 & 25 & 0 & $2.49 \pm 0.22$ & $3.85 \pm 1.02$ & $6.37 \pm 0.03$ & $0.24 \pm 0.09$ & $24.0 \pm 2.8$ & 2.32 & $-108.8 \pm 2.2$ & $19.1 \pm 2.0$ \\ 
20 & 35 & 4 & 0 & $2.54 \pm 0.20$ & $3.84 \pm 0.90$ & $6.35 \pm 0.03$ & $0.23 \pm 0.08$ & $22.2 \pm 2.6$ & 2.47 & $-127.6 \pm 2.1$ & $21.9 \pm 2.1$ \\ 
20 & 40 & 0 & 0 & $2.55 \pm 0.19$ & $3.83 \pm 0.53$ & $6.37 \pm 0.03$ & $0.25 \pm 0.06$ & $20.8 \pm 2.4$ & 2.51 & $-147.4 \pm 2.0$ & $23.4 \pm 2.1$ \\ 
20 & 50 & 0 & 0 & $2.61 \pm 0.18$ & $3.63 \pm 0.29$ & $6.36 \pm 0.02$ & $0.24 \pm 0.05$ & $21.3 \pm 1.6$ & 2.69 & $-184.9 \pm 2.7$ & $30.2 \pm 2.4$ \\ 
\hline

\end{tabular}
\end{table*}

\section{Conclusions}\label{sec:section5}

In this work, we simulated mass measurements of the young planet K2-33b, using synthetic RV time-series assuming different sampling schemes, levels of white noise and semi-amplitudes of the planet signal. We found that 35/50 SPIRou visits spread on a 3-month visibility window are enough to reliably detect planet signatures of 10/5~\ms\ at precisions $>3\sigma$ at a white noise level $\la$5~\ms. Conversely, 30 visits over the same time span often yield synthetic data sets that can be inconclusive about the presence of the planet or lead to large errors in the estimated mass.


As suggested in \citetalias{mann2016}, the substantial decrease in the stellar activity RV signal from visible wavelengths to the nIR makes K2-33 ideally suited for RV observations with high-precision nIR velocimeters like SPIRou. Our simulations demonstrate that the RV signature of the close-in Neptune-like planet K2-33b can be detected at a 3$\sigma$ level for a planet mass of $\ga$0.5~\mn, provided that the observational sampling is dense enough.

K2-33b being significantly younger than all the planets with well-measured masses and radii, its nature is still unclear. Measuring the planet mass and hence probing its inner structure and composition will help unveiling its nature as well as constraining the evolution of its bulk density throughout the formation stage \citep[see the review from][]{baruteau2016}.

Consistently with \citetalias{klein2019}, this study demonstrates the crucial need for RV follow-ups of stars hosting transiting planets to densely cover both the planet orbital period and stellar rotation cycles, on timescales that are of the same order of the one on which stellar activity changes. Stars with activity features evolving on timescales of tens of days, and rotation periods of only a few days, are therefore the most suited targets for RV follow-ups with high-precision velocimeters, as the mass of their transiting companions can be accurately measured in a single $\sim$90-d window. NIR RV observations with sampling schemes similar to that we propose, as those being currently carried out in the framework of the SPIRou Legacy Survey, are thus likely to provide soon reliable mass measurements of transiting PMS planets such as K2-33 and V1298 Tau \citep{david2016,david2019,david2019b}.

\section*{Acknowledgements}

This project was funded by the European Research Council (ERC) under the H2020 research \& innovation programme (grant agreements \#740651 NewWorlds). The authors would like to thank the referee for valuable comments and suggestions which helped improving the manuscript.




\bibliographystyle{mnras}
\bibliography{bibliography} 



\bsp	
\label{lastpage}
\end{document}